# The uncovering of hidden structures by Latent Semantic Analysis

Juan C.Valle-Lisboa[*][!] & Eduardo Mizraji[*]

Sección Biofísica, Facultad de Ciencias, Universidad de la República, Iguá 4225, Montevideo 11400, Uruguay.

**Categories and subject descriptors**: H.3.3[Information storage and retrieval] Information search and retrieval – Retrieval models, Clustering  H.3.1[Information storage and retrieval] Content analysis and Indexing – Indexing methods.

[*] Facultad de Ciencias,
  Iguá 4225, Montevideo 11400,
  Uruguay.

[!] Corresponding author.
  Phone: 5982-5258618 to 23 ext. 7139.
  Fax: 5982-5258617.
  Email: juancvl@fcien.edu.uy

# 0. Abstract

Latent Semantic Analysis (LSA) is a well known method for information retrieval. It has also been applied as a model of cognitive processing and word-meaning acquisition. This dual importance of LSA derives from its capacity to modulate the meaning of words by contexts, dealing successfully with polysemy and synonymy. The underlying reasons that make the method work are not clear enough. We propose that the method works because it detects an underlying block structure (the blocks corresponding to topics) in the term by document matrix. In real cases this block structure is hidden because of perturbations. We propose that the correct explanation for LSA must be searched in the structure of singular vectors rather than in the profile of singular values. Using Perron-Frobenius theory we show that the presence of disjoint blocks of documents is marked by sign-homogeneous entries in the vectors corresponding to the documents of one block and zeros elsewhere. In the case of nearly disjoint blocks, perturbation theory shows that if the perturbations are small the zeros in the leading vectors are replaced by small numbers (pseudo-zeros). Since the singular values of each block might be very different in magnitude, their order does not mirror the order of blocks. When the norms of the blocks are similar, LSA works fine, but we propose that when the topics have different sizes, the usual procedure of selecting the first k singular triplets (k being the number of blocks) should be replaced by a method that selects the perturbed Perron vectors for each block.

# 1. Introduction

One of the most important methods for semantic retrieval of information is Latent Semantic Analysis (LSA, also called Latent Semantic Indexing, LSI) [Deerwester et al. 1990]. In this method a word-document matrix is approximated using a limited (but still high) number of orthogonal factors. This conceptually simple procedure allows for the search of relevant information greatly avoiding the problems of polysemy and synonymy.

The technique has been successfully applied to cognitive psychology and mimics several psychological results (see [Dumais 2003] for a review). For instance, Landauer and Dumais [Landauer and Dumais 1997] have applied LSA to word learning, addressing one aspect of the important "Plato's problem" of how we can learn a great amount of information from few data, a general problem that has been, and still is, central in linguistic debates (see for example [Chomsky 1985] and [Lewis and Elman 2002]).

Latent Semantic Analysis is based on approximating the word-document matrix with the singular value decomposition (SVD). Our interest in the technique stems from the finding that SVD arises naturally from certain types of context-dependent memory models [Pomi and Mizraji 1999]. Given the link between cognition and LSA, we have a strong interest in understanding the detailed functioning of LSA.

The diverse applications of LSA arise from its ability to uncover the hidden topic structure in the input matrix, building a semantic space. The subsequent projection of words and documents over this semantic space allows the comparison of portions of text of various sizes.

The superficial simplicity of the recipe for building a Latent Semantic representation hides the understanding of why the method works. Theoretically, the reasons for the success of the method and its capacity to deal with polysemy and synonymy are far from clear [Hoffmann, 1999; Langville and Meyer, 2004]. The discussion of this topic is usually framed in the well known theorem of Schmidt – Eckart and Young (usually called Eckart and Young theorem, see below and [Stewart 1993]), which states that the truncated SVD (which is the matrix used in LSA) is the best rank-k approximation to the original matrix. Nevertheless, as has been pointed out by other authors [Papadimitriou et al 2000; Ando and Lee 2001] this theorem can only be part of the explanation of LSA's power. In the last few years many analyses of LSA have been published, mainly based on the application of invariant subspace perturbation theory [Ando and Lee 2001; Azar et al 2001; Papadimitriou et al 2000; Zha and Zhang 1999]. We discuss the connections of these previous approaches to our work in the following sections, particularly in section 8. Our results complement those of other authors but are based on a different approach. In this paper we study LSA basing our discussion on Perron-Frobenius Theory of Nonnegative matrices (described in [Meyer 2000], chapter 8). Although we cannot give a complete answer to the question of what makes LSA work, we think that our approach can help bridge the gap between the practical and theoretical aspects of the method.

In the following section (section 2) we give a short summary of LSA. To understand how LSA works and its possible pitfalls, we introduce an artificial example with clearly separated topics in section 3. This example allows us to present an ideal topic structure with no words shared by different topics and single-topic documents (i.e. documents relevant to only one topic), and to study how LSA deals with this topic structure. We show that no simple recipe about the profile of singular values can help in selecting the appropriate singular values and vectors to represent documents. We then look into the singular vectors for clues about topic structure (Section 4). The application of the well-known Theory of Nonnegative Matrices implies that in the simple case of unconnected topics each topic will be represented by a right singular vector (and its associated left vector and singular value) with sign homogeneous entries corresponding to documents relevant to the topic and zeros elsewhere. Conversely any singular vector with this structure indicates that there are at least two disconnected topics (Section 5). Aided by our simple example, *we show that if the database contains k topics, the singular values associated with them are not necessarily the first k singular values*. For instance, in our example with three topics, the first, the second and the fifth singular values are the leading ones associated with each topic. *We argue that if the hidden Perron vectors can be detected, just retaining these vectors is enough to capture the topic structure*.

To apply these ideas to realistic examples we introduce Perturbation Theory in Section 6 aiming to consider realistic examples as perturbed ideal collections. We show that if the underlying topics are associated with well separated singular values, the zeros of the ideal cases map in small entries in the relevant singular vectors. In Section 7 we apply the ideas developed throughout the paper to improve the working of LSA in a moderate size example, namely, a subset of the OHSU-MED database.

## 2. A short review of the basic LSA method.

The method starts first by writing $A_{mxn}$, the matrix of word by document co-occurrence which takes the form

$$A = [\mathbf{d}_1 \vdots \mathbf{d}_2 \vdots \mathbf{d}_3 \vdots \ldots \vdots \mathbf{d}_n]; \quad \mathbf{d}_i = \begin{bmatrix} a_{1i} \\ a_{2i} \\ \vdots \\ a_{mi} \end{bmatrix}, \quad (1)$$

where the $\mathbf{d}_i$s are document vectors having in their coordinates $j$ the number of times the word $j$ appears in document $i$.

An optional step is to weight each entry in the matrix using appropriate weighting schemes which can greatly enhance the performance of the method. We will not show these weighting schemes here. For a thorough review see [Dumais 1991].

Once $A$ is formed (and possibly weighted) the singular value decomposition of $A$ is computed. It is a well-known theorem of linear algebra [Meyer 2000] that every matrix $A$ has such a decomposition. In general the singular value decomposition of $A$ is,

$$A = U_{mxr} \Sigma_{rxr} V_{nxr}^T = \sum_{s=1}^{r} \sigma_s \mathbf{u}_s \mathbf{v}_s^T \quad ,$$

where $r$ is the rank of $A$, $\sigma_s$ is the $s^{th}$ the singular value of $A$, and $\mathbf{u}_s$ and $\mathbf{v}_s$ are the corresponding left and right singular vectors. (In this case we omit the singular vectors associated with singular values equal to zero i.e. the vectors that form a basis for the null space of $A$ or $A^T$). The columns of the matrix $U$ constitute an orthonormal basis for the range of matrix $A$ and the columns of $V$ constitute an orthonormal basis for the range of $A^T$. Notice that these bases are not independent.

In LSA the procedure is to construct a matrix $A_k$, using the truncated SVD, which is an approximation of rank $= k < r$ to the original matrix, that is,

$$A_k = \sum_{s=1}^{k} \sigma_s \mathbf{u}_s \mathbf{v}_s^T \quad . \tag{2}$$

LSA has been primarily used as an information retrieval method [Deerwester et al. 1990], where a query vector can be compared to the reconstituted matrix or mapped to the reduced space and compared with the representation in this space of the documents. After these initial applications there has been a growing interest in the technique both from a technological and from a psychological point of view (reviewed by [Dumais 2003]). Yet, as is recognized by some authors, the reasons that make the method work are obscure. It is generally agreed that an important step to understanding the functioning of the method is the well known theorem often attributed to Eckart & Young (but which has been previously established by Schmidt, see [Stewart 1993]):

*Theorem 1*: Given a matrix *A* of rank r with singular values σ$_1$,...,σ$_r$ for any k <r, and let *A*$_k$ be defined as in equation 2, then

$$\min_{\text{rank}(B)=k} \|A-B\|_F^2 = \|A-B_k\|_F^2 = \sigma_{k+1}^2 + \ldots + \sigma_r^2$$

This shows that of all rank-k (and lower rank) matrices, the reduced matrix A$_k$ is the best approximation (in the Frobenius norm) to the original matrix. Clearly, this optimality is needed in order to reproduce the original data, but it cannot explain why and where we need to cut the description and how many factors we need to retain.

Generally speaking, the qualitative properties of the model are said to be based on a reduction of linguistic noise [Deerwester et al 1990].

**On topics and clusters**.

Although we will not explicitly state a model of topic structure, in the following, we implicitly assume that documents which are relevant to the same topic should be clustered together when low-rank approximations are used. The idea is that a topic is defined in terms of a set of "marker" words, and documents relevant to that topic have some of these words or at least share words with other documents relevant to the topic. In the traditional vector space model only those documents that share words are considered to be similar. In our conception if there is a chain of words connecting two documents they are similar, and the strength of this similarity is dependent upon the word pathways connecting the two documents. If the sets of words that define each topic are disjoint and documents are relevant to a single topic, then each topic can be mapped in a neat cluster of documents. We want to see in which cases LSA clusters documents belonging to the same topic and how it deals with perturbations. One of the most important perturbations are those made by polysemous words, since a word with different meanings can be one of the "topic markers" we were referring to above, and consequently not be a small perturbation.

## 3. An example to illustrate the method.

We show here an example of the application of LSA to a small database. Table 1 shows the titles of some articles that we use in our courses in our Uruguayan University. As is customary in LSA (e.g. see [Berry et al. 1995]) we only retain the words that appear more than twice in the database. Moreover to reduce overlapping we stop-listed words (*a, and, by, for, from, in, non, of, on, the*). Likewise, we transformed the following words to their singular form: *features, networks, memories,* but we used no stemming methods, whose application to LSA can increase or decrease performance [Dumais 1991].

In this small example there are clearly three disjoint topics, although there are words that are shared by all the topics, both generic (*probabilistic, model, theory, based*) and polysemous (*kinetic* which is used both to mean the kinetic theory on matter and chemical, macroscopic kinetic phenomena).

In order to enforce the existence of disjoint clusters we removed the words *probabilistic, theory, model, kinetic, based*. We call the resulting database "pruned example 1" (PE1). The construction of this database ensures the existence of three blocks: one related to Brownian motion, another related to allosteric proteins and a third one related to neural networks.

| | | |
|---|---|---|
| | 1 | On the <u>movement</u> of small particles suspended in a stationary liquid demanded by the <u>molecular</u>-kinetic theory of heat. |
| | 2 | On the theory of <u>Brownian</u> <u>movement</u>. |
| | 3 | New determination of <u>molecular</u> dimensions. |
| | 4 | Theoretical observations on the <u>Brownian</u> <u>motion</u>. |
| | 5 | Elementary theory of the <u>Brownian</u> <u>motion</u>. |
| | | |
| | 6 | On the <u>nature</u> of <u>allosteric</u> <u>transitions</u>: A plausible model. |
| | 7 | A probabilistic approach to cooperativity of <u>ligand</u> <u>binding</u> by a polyvalent molecule. |
| | 8 | Kinetics of the <u>allosteric</u> <u>interactions</u> of phosphofructokinase from <u>Escherichia</u> <u>coli</u>. |
| | 9 | <u>Allosteric</u> <u>receptors</u> after 30 years. |
| | 10 | A kinetic mechanism for nicotinic acetylcholine <u>receptors</u> based on multiple <u>allosteric</u> <u>transitions</u>. |
| | 11 | <u>Allosteric</u> <u>interactions</u> interpreted in terms of quaternary structure. |
| | 12 | On the <u>nature</u> of <u>allosteric</u> <u>transitions</u>; implications of non exclusive <u>ligand</u> <u>binding</u>. |
| | 13 | <u>Allosteric</u> <u>Transition</u> and Substrate Binding Are Entropy-Driven in Glucosamine-6-Phosphate Deaminase from <u>Escherichia</u> <u>coli</u>. |
| | | |
| | 14 | The perceptron: a probabilistic model for information <u>storage</u> and <u>organization</u> in the brain. |
| | 15 | Correlation <u>matrix</u> **memories**. |
| | 16 | A simple <u>neural</u> <u>network</u> generating an interactive <u>memory</u>. |
| | 17 | A possible <u>organization</u> of animal <u>memory</u> and <u>learning</u>. |
| | 18 | Adaptive pattern classification and universal recoding I :Parallel development and coding of <u>neural</u> <u>feature</u> detectors. |
| | 19 | <u>Neural</u> theory of association and concept <u>formation</u>. |
| | 20 | <u>Neural</u> **networks** and physical systems with emergent collective computational abilities. |
| | 21 | Non-holographic associative <u>memory</u>. |
| | 22 | <u>Learning</u> representations by backpropagating errors. |
| | 23 | Theory of categorization based on distributed <u>memory</u> <u>storage</u>. |
| | 24 | Self-organized <u>formation</u> of topologically correct <u>feature</u> maps. |
| | 25 | Distinctive **features**, categorical perception and probability <u>learning</u>: some applications of a <u>neural</u> model. |

**Table 1**: The original 25 titles used to illustrate the different aspects treated in the present article. In order to construct Pruned example 1(PE1) we retained the underlined words and transformed the bold words in singular form. Notice that certain words : Database of titles after the transformations described in text in order to artificially create three clear blocks; the first one (documents 1 to 5) is related to Brownian motion; the second one (documents 6 to 13) with titles from the field of allosteric proteins; the third one (14 to 25) is related to neural networks. Underlined words are retained for LSA.

In figure 1 we show the word-document matrix A of Pruned Example 1, and its profile of singular values.

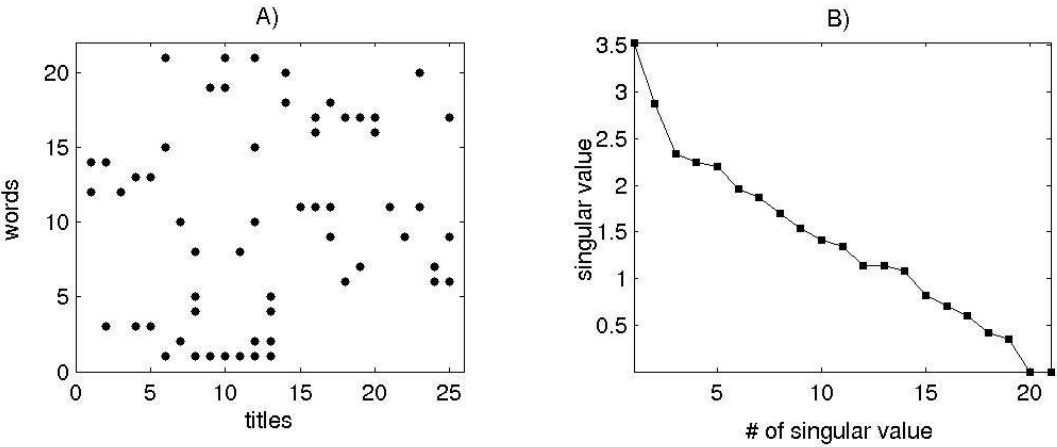

**Figure 1**: A) Word-document matrix for pruned example 1 (the abscissas index the documents and the ordinates the terms). The black squares are the only nonzero entries. B) Profile of singular values corresponding to the matrix in A.

To obtain a useful semantic representation it has to be decided how many singular values and vectors have to be retained. Following usual procedures in Multivariate Statistics literature we could cut the profile of singular values where there is a noticeable drop off, a procedure related to Cattell's scree test for Principal Components Analysis [Cattel 1966]. We show below, that for our simple data it is not a simple decision to apply these criteria.

Notice that in our artificial example we can *a priori* assign documents to the three different topics, which in term of words is the same as assigning the words present in each topic to one of three partitions. To show the difficulties arising when trying to define the number of singular values to be retained, we compare the profile of singular values after shuffling the words across all the documents. We constructed different word-document matrices by selecting words in table 1 and placing them randomly in one of the documents, with the restrictions that each documents has at least one word and the total frequency of each word is preserved. The main point we wish to make is that if singular values show a clear cutting point because of similarity structure, we expect a qualitatively different profile of singular values after shuffling.

Following Montemurro and Zanette [Montemurro and Zanette 2002] we measure the disorder of words after shuffling with the entropy for each word x, S(x), in the three natural partitions that arise in this simple case as follows:

$$S(x) = -\frac{1}{\ln P} \sum_{j=1}^{P} p_j(x) \ln p_j(x) \quad , \quad (3)$$

where $p_j(x)$ is the probability that, given that the word x was chosen at random, it comes from partition $j$, and is estimated as,

$$p_j(x) = \frac{n_i(x)/N_j}{\sum_{i=1}^{P} n_i(x)/N_i} \quad ,$$

and in both expressions P is the number of partitions $N_j$ is the total number of words in each partition and $n_i(x)$ is the number of times that word $x$ is in partition $i$. Notice that the entropy (eq 3) is normalized and goes between 0 and 1, the former meaning that the word is to be found in only one partition and the latter meaning that the word is uniformly distributed among the partitions.

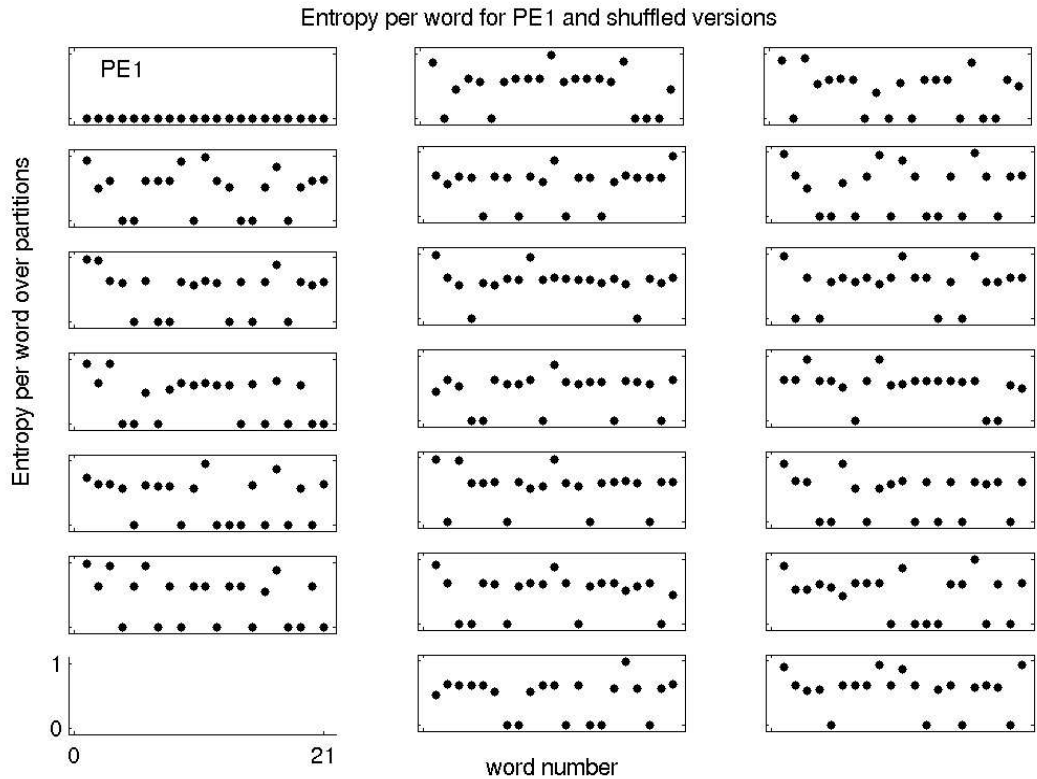

**Figure 2** Entropy of the words (equation 3) calculated using the relevant blocks (topics in table II) as partitions, for the example PE1. The upper left graph corresponds to the original case; the rest are shuffled versions, but retaining the definition of partitions. In each window the entropy per word is displayed; the y-axes go from -0.1 to 1.1 and the x-axes from -0.5 to 22, although entropies go from 0 to 1 and there are 21 words.

In figure 2 the entropy of each word is shown for the case of Pruned Example 1 (first graph) and 19 associated shuffled cases. In the first upper-leftmost window (the original PE1 case) the entropies are all zero as desired. In all the other cases there are increases of entropies of most words, a signature that the underlying structure is destroyed. We set to see whether these disordered states map somehow in the profile of singular values. To our surprise these profiles (shown in figure 3) seem to vary little and not in a regular fashion with entropy. A lowering of the first singular value of the scrambled databases compared to the ordered case seems to repeat itself, though in other cases (not shown) this difference is marginal.

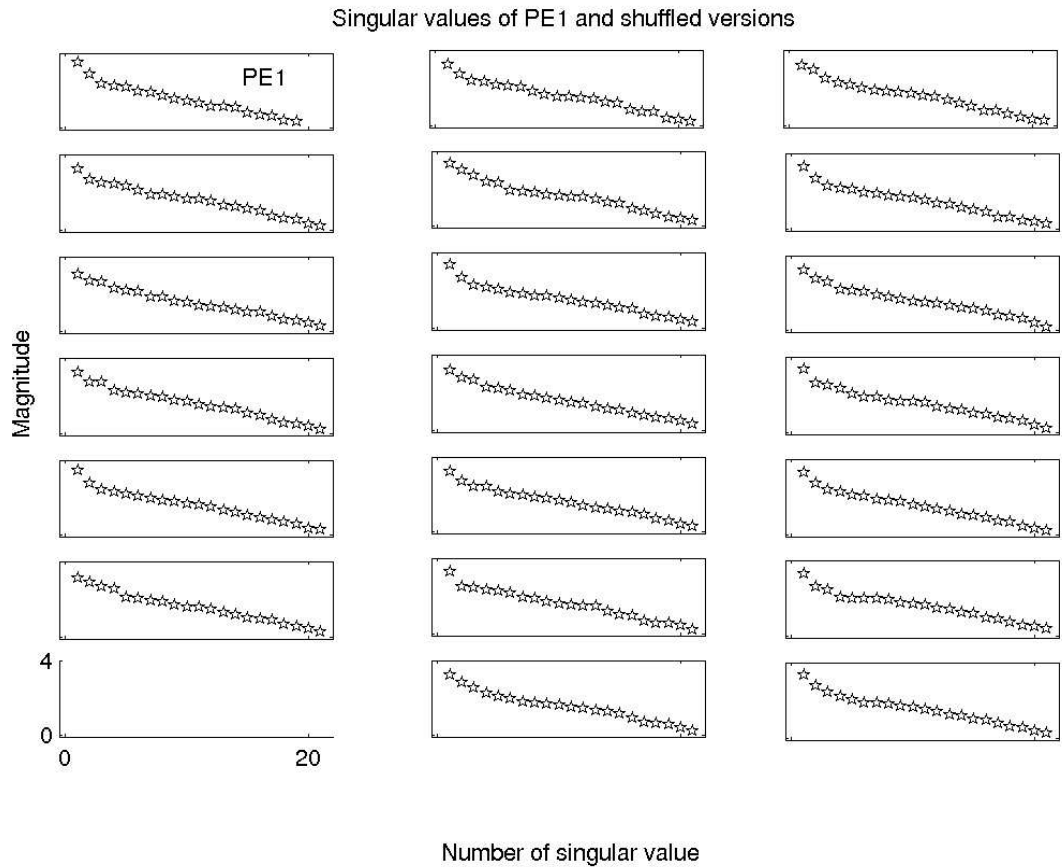

**Figure 3**: Profile of singular values for the 20 cases of figure 2 where the upper-left case is PE1 and the other 19 correspond to precisely the same shuffled examples of figure 2. The y-axes all go from -0.1 to 4.1 and the x-axes go from -0.1 to 22.

There are no obvious discontinuities in the profile of singular values where the decision to cut is relevant for obtaining the correct latent semantic space. It can be argued that in some of the scrambled cases there are new randomly arranged clusters of documents that do not coincide with the chosen partitions, but this is not a general phenomena and yet the increase in entropy with no alteration of singular values is always observed. Knowing in advance the number of blocks is not enough to select the number of factors. If we select just the first three singular values (and their associated vectors), the information of the Brownian motion block is lost and no query would recall documents of this topic. We describe two ways to circumvent this problem:

## a) The traditional LSA approach

One alternative is to include more factors until documents of Brownian motion appear. In figure 4 we show that when a truncated SVD of dimension 5 is used, all documents can be recalled and the cosines between documents relevant to the same topic are non-zero whereas documents of different topics are orthogonal. Yet, the presence of negative cosines is difficult to interpret. For instance in figure 4, document 18 is positively correlated to document 16 (cosine=0.76), and document 16 is positively correlated to document 14 (cosine=0.36), yet documents 14 and 18 are negatively correlated (cosine=-0.34). Although this effect is surely due to the small size of the example, whenever negative correlations appear it is not simple to give an interpretation in terms of similarity.

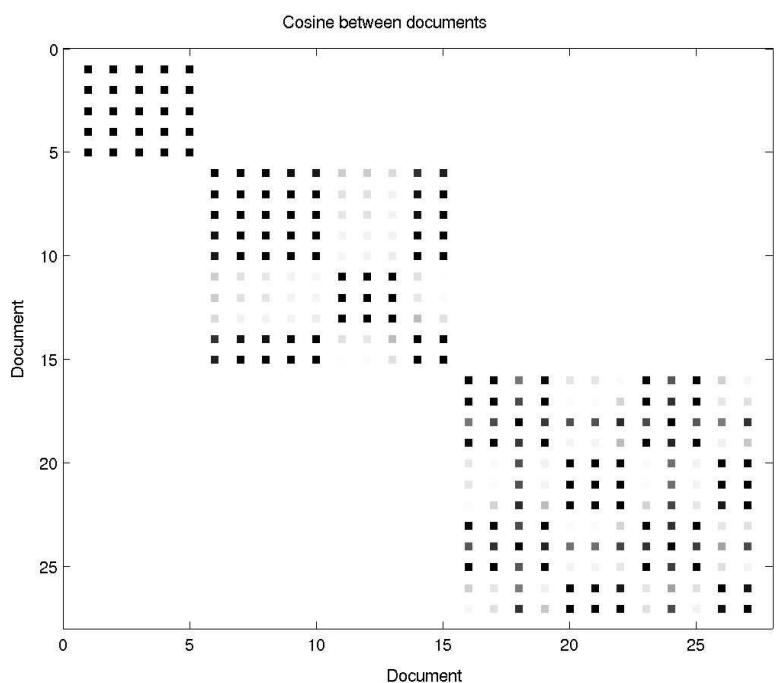

**Figure 4**: Document-Document Correlation Matrix: Cosines between documents when the example PE1 is represented by the first 5 singular triplets. In black cosines equal to one, in gray we mark those cases where cosines are lower than 1 or negative and in white cases where the cosines are 0.

### b) A different approach: select non-consecutive singular triplets

Another strategy would be to take the first, the second and the fifth singular values (and their associated vectors). This selection makes documents from the same topic show positive correlations. This is shown in figure 5.

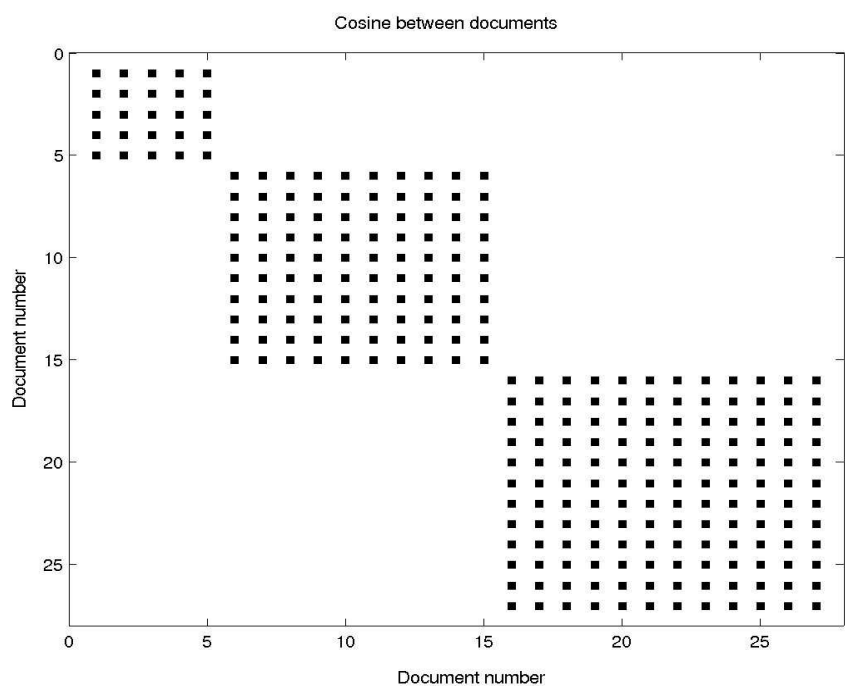

**Figure 5**: Document correlation matrix (cosines) between documents for the example of table II (PE1), when the singular values and vectors retained are the first, the second and the fifth. The black squares correspond to cosines=1; all the other cosines are = 0.

---

As can be seen, there is no obvious way of deducing from the profile of singular values how many singular triplets should be retained to detect the latent semantic structure. We show next that the relevant information might come from the singular vectors themselves.

## 4. The structure of the correlation matrix.

We then ask what can be said about the correlations between documents and their relation to singular values and vectors. We will only use dot product correlations which are more amenable to algebraic treatment. Even though we will mainly refer to correlation between documents, given that the technique simultaneously groups words and documents the same results can be applied to correlation between words.

First note that the product $A^TA$ gives the correlation between documents and can be written

$$C = A^T A = \sum_{j=1}^{n} \sum_{k=1}^{n} \langle \mathbf{d}_j, \mathbf{d}_k \rangle \mathbf{e}_j \mathbf{e}_k^T, \qquad (4)$$

where $e_j$ and $e_k$ are vectors of the canonical base of $R^{n \times 1}$, having zeros in every position except in the coordinate i and j respectively (i.e. they are the $i^{th}$ and $j^{th}$ columns of the identity matrix). Notice that in the case of orthogonal documents, the rank of matrix *A* is *n* and only the terms where *j=k* are retained in equation 4. In this case, the singular values are the Euclidean norms of document vectors, showing that the ordering of singular values is strongly dependent on the "size" of the documents. This orthogonal documents' example is an extreme one, but the simple conclusion carries on to more general cases.

It cannot be concluded that singular values are ordered, in the sense that the first k singular values belong to k different clusters of documents. Cases like those of figure 4, where the third topic is only represented if we include at least five singular vectors, are to be expected whenever the different topics are unequally represented in the database.

The elements of the correlation matrix $C_{nxn} = [\langle \mathbf{d}_i, \mathbf{d}_j \rangle]$, can be written in terms of the singular value decomposition, since,

$$C = A^T A = V \Sigma^2 V^T, \qquad (5)$$

these coordinates are,

$$[c_{ij}]_{nxn} = \langle \mathbf{d}_i, \mathbf{d}_j \rangle = \sum_{h=1}^{h=r} \sigma_h^2 v_h(i) v_h(j), \qquad (6)$$

where $v_h(i)$ and $v_h(j)$ are components *i* and *j* of the h[th] right singular vector. As these considerations show, there are interesting features about document – document correlations that depend not only on singular values but also on the components of the singular vectors. We turn now to the application of the Perron-Frobenius theory of nonnegative matrices in order to study what to expect from those components.

## 5. Application of Perron-Frobenius Theory

Searching for traces of the structure in the example of figure 4 in the singular vectors v (the columns of the V matrix) it can be seen that each block is represented in the sign-homogeneous nonzero entries of singular vectors (see table 2 below). Given that in Pruned Example 1 there are disjoint blocks, the vectors have zeros in the other positions. We analyze the reasons for these behaviors using Perron-Frobenius Theory of nonnegative matrices.

For completeness we review (without demonstrations) the basic theorems of Perron-Frobenius theory (see [Meyer 2000] chapter 8).

<u>*Definition 1*</u> : A matrix **D** is said to be <u>reducible</u> if there exists a permutation matrix **P** such that

$$P^T DP = \begin{pmatrix} X & Y \\ 0 & Z \end{pmatrix}$$, where **X** and **Z** are square matrices.

In any other case **D** is an <u>irreducible</u> matrix.

<u>*Definition 2*</u>: A matrix which can be brought by permutations to have square blocks in its diagonal and zeros everywhere else is said to be a decomposable matrix (or a block-diagonal matrix).

It is important to note that any nonnegative square matrix (of n rows by n columns) has an associated graph of *n* nodes and the nonzero elements *ij* of the matrix represent links between nodes *i* and *j*. It can be demonstrated that a matrix is irreducible if and only if there is a sequence edges leading from one arbitrary node to every other node of the underlying graph; in those cases the graph is said to be strongly connected.

We present a reduced version of Perron-Frobenius Theorem (Theorem 2).

*Theorem 2*: Let $D \geq 0$, $D \in R^{n \times n}$ be an irreducible matrix. Then:

    a- The spectral radius r of the matrix $D$ belongs to the spectrum of $D$ and is a positive eigenvalue with algebraic multiplicity = 1.

    b- There exists an eigenvector $v > 0$ such that $Dv = rv$.

    c- The unique vector defined by $Dp = rp$, $p > 0$, and $\|p\|_1 = 1$ is called the Perron vector (and the pair $(r,p)$ is the Perron Pair. There are no other nonnegative eigenvectors for $D$ except for positive multiples of $p$, regardless of the eigenvalue.

*Theorem 3*: Given a nonnegative matrix $D$, there is a positive integer m such that $D^m > 0$ if and only if $D$ is irreducible and the eigenvalue r is the only one in the spectral circle.

In our case $D = A^T A$, is by definition nonnegative and symmetric as is $W = AA^T$. This simplifies the meaning of the theorems since no eigenvalues are complex. Moreover, as we show below these properties allow us to deduce irreducibility from a positive eigenvector.

Theorem 2 has a nice graph theoretic interpretation; if every entry in the matrix is interpreted as the weight of a directed arrow connecting two nodes in a graph of n nodes, the theorem states that when all nodes are connected by a path of documents the first eigenvector of matrix $D$ will be strictly positive. Thus, we expect the first singular vector of matrix $A$ to be positive when the documents are connected. The contrapositive of theorem 2c) implies that whenever we find zeros in the first singular vector we expect some documents to be disconnected from others.

It is not generally valid that a positive eigenvector and a nonnegative matrix ensure that all the documents are connected. Yet, since matrix $D$ is symmetric, the following lemma shows that in this case a positive eigenvector implies that all documents are connected.

*Lemma 1*: Let **D** be the nonnegative and symmetric matrix $D=A^T A$ and let $\sigma_1$ be the highest eigenvalue of **D**. If the $\sigma_1$ has algebraic multiplicity = 1 and the eigenvector **v** associated to $\sigma_1$ is positive, then **D** is irreducible (i.e. there is a path connecting every document to every other document).

*Proof*: Let r be the rank of matrix *A*. The validity for a matrix of r=1 is trivial. For r ≥ 2, for a positive integer k we have, following equations 5 and 6 and recalling that $V^T V = I$,

$$D^k = (A^T A)^k = V \Sigma^{2k} V^T,$$

then each entry $\delta_{ij}^{(k)}$ of the matrix $D^k$ can be written as,

$$\delta_{ij}^{(k)} = \sum_{s=1}^{r} \sigma_s^{2k} v_s(i) v_s(j) .$$

Since all the components of $v_1$ are positive, every coordinate of $D^k$ has a positive contribution from the first term. In the worst case, every other contribution to $\delta_{ij}$ is negative and then,

$$\delta_{ij}^{(k)} > \sigma_1^{2k} v_1(i) v_1(j) - (r-1) \sigma_2^{2k} .$$

We can define a number $k_{ij}$ which represents the minimal exponent k that makes $\delta_{ij}^{(k)}$ positive. The latter equation shows that $k_{ij}$ is bounded by,

$$k_{ij} \geq \frac{\log\left(\frac{r-1}{v_1(i) v_1(j)}\right)}{\log\left(\frac{\sigma_1}{\sigma_2}\right)^2} .$$

Since r ≥ 2, $\sigma_1 > \sigma_2$ and $v_1$ is a normalized positive vector, taking the $k = \max_{ij}\{k_{ij}\}$ ensures that every entry in the matrix $D^k$ is positive, for finite k. According to theorem 3 this implies that the matrix is irreducible .

Notice that if the algebraic multiplicity of the first eigenvalue is higher than one, then by theorem 2a, the matrix cannot be irreducible. This can happen if there are two separate blocks in the database sharing their first singular value.

Lemma 1 admits the following interpretation recalling that the square matrix $D$ (*nxn*) represents a graph of n nodes. A matrix is irreducible if every node can be reached from every other node. The existence of a positive Perron pair implies that there is at least a path of undirected edges connecting any two nodes. No general statement about reducibility can be said if the matrix is not symmetric for there can be nodes acting as "sinks". Given that $A^TA$ is symmetric, if there is a path from node *i* to node *j* then there is a path from node *j* to *i* and this ensures the irreducibility.

Thus, if one performs the singular value decomposition of the word-document matrix $A$, when the first singular vector $v_1$ is positive, the documents are connected and the product of the coordinates provides the correlation between them. Moreover, in this case, the theorem 2c. ensures that the remaining terms of equation 2 will have mixed negative and positive coordinates deteriorating some of the correlations and enhancing others. If the first singular vector has zeros then the matrix $D$ is reducible. Finally we have the following property (the proof is omitted):

*Property 1*: Let $D_{nxn}$ be a symmetric matrix. If $D$ is reducible then it is decomposable.

Property 1 together with Theorem 2 implies that whenever we find true zeros in the first singular vector of matrix $A$ there will be at least two blocks of uncorrelated documents. Since those blocks arise as the product of unconnected regions of the original matrix $A$, we have different topics in the document database that could be handled separately. Each cluster will be defined by the nonzero entries of a sign-homogeneous singular vector. We will call these vectors the leading vector of each block. The number of clusters is equal to the number of leading vectors, as implied by Theorem 2c.

The theorems thus far presented allow us to explain pruned example 1. In table 2 we present the first five singular values and their associated right singular vectors.

|     | v1     | v2     | v3      | v4      | v5     |
|-----|--------|--------|---------|---------|--------|
| d1  | 0      | 0      | 0       | 0       | 0.1924 |
| d2  | 0      | 0      | 0       | 0       | 0.4927 |
| d3  | 0      | 0      | 0       | 0       | 0.0503 |
| d4  | 0      | 0      | 0       | 0       | 0.5991 |
| d5  | 0      | 0      | 0       | 0       | 0.5991 |
| d6  | 0.3751 | 0      | 0       | -0.3012 | 0      |
| d7  | 0.1378 | 0      | 0       | -0.2071 | 0      |
| d8  | 0.3846 | 0      | 0       | 0.6176  | 0      |
| d9  | 0.2531 | 0      | 0       | 0.0248  | 0      |
| d10 | 0.3524 | 0      | 0       | -0.1693 | 0      |
| d11 | 0.2559 | 0      | 0       | 0.2194  | 0      |
| d12 | 0.5129 | 0      | 0       | -0.5082 | 0      |
| d13 | 0.4190 | 0      | 0       | 0.3864  | 0      |
| d14 | 0      | 0.0805 | -0.2612 | 0       | 0      |
| d15 | 0      | 0.1588 | -0.2966 | 0       | 0      |
| d16 | 0      | 0.4907 | -0.1222 | 0       | 0      |
| d17 | 0      | 0.3124 | -0.4791 | 0       | 0      |
| d18 | 0      | 0.3514 | 0.3179  | 0       | 0      |
| d19 | 0      | 0.2884 | 0.2570  | 0       | 0      |
| d20 | 0      | 0.3319 | 0.1744  | 0       | 0      |
| d21 | 0      | 0.1588 | -0.2966 | 0       | 0      |
| d22 | 0      | 0.1060 | -0.0467 | 0       | 0      |
| d23 | 0      | 0.1918 | -0.4219 | 0       | 0      |
| d24 | 0      | 0.1752 | 0.2452  | 0       | 0      |
| d25 | 0      | 0.4574 | 0.2712  | 0       | 0      |

**Table 2:** The first five right singular vectors of pruned example 1 (PE1). The first, the second and the fifth singular vectors are the leaders of each block. (The first five eigenvalues of $A^TA$ are 12.4815, 8.2633, 5.4501, 5.0428, and 4.8224, i.e the first 5 singular values are 3.5329, 2.8746, 2.3345, 2.2456, 2.1960 ).

---

The first singular vector shows that matrix $A$ has an irreducible block consisting of documents 6 to 13; the second vector shows the presence of another block of documents going from 14 to 25. In order to include documents 1 to 5 we need to reach the fifth singular vector. But notice that, according to the Perron-Frobenius theory, the other vectors related to documents 6 to 25 (the third and fourth vector) should have mixed signs in their nonzero entries. By equation (6) the correlations within the blocks will be reduced if these vectors are retained, and can be made negative. These negative correlations are difficult to interprete and might deteriorate performance. We discuss this problem when applying our ideas to OHSU-MED. In any case, the lowest rank approximation (but not necessarily the best in the Frobenius norm sense) to the matrix associated with Pruned Example 1 is obtained by retaining all the nonnegative singular vectors something that can be achieved in a simple way, solving, for this case, the dimensions' selection problem.

We are not addressing here the problem of distinguishing the zeros from small entries in the singular vector (which might be important if one considers the output of numerical methods). Nevertheless, a "true zero" in a singular vector whose coordinates are sign homogeneous is paralleled by a coordinate different from zero in other vectors with homogeneous signs in their nonzero entries. Moreover, as we show below, small entries might signal weak links that can be ignored in a first approximation.

The results thus far presented can be applied to words. If instead of matrix $\boldsymbol{D}=\boldsymbol{A}^T\boldsymbol{A}$ we consider the matrix $\boldsymbol{W}=\boldsymbol{A}\boldsymbol{A}^T$, we are referring to word-word correlations. The analogous to equation (6) has now the entries of the left singular vectors. Perron Frobenius theory guarantees that if the matrix W is reducible there will be one nonnegative left singular vector for each topic that has nonzero entries only in those entries corresponding to words belonging to that topic. Reciprocally, the positive entries of a left nonnegative singular vector correspond to related words.

## 6. Real cases as perturbations of decomposable matrices.

The main results of the previous section are that for each pure topic there is a corresponding nonnegative right singular vector. Moreover, the associated singular values are not successive, that is, the singular value corresponding to the $k^{th}$ topic is not necessarily the $k^{th}$ singular value.

In real applications, separate blocks might appear less frequently than cases with interconnected blocks. Yet, we would like to see if in some of these cases there are patterns that justify a separation of the matrix in blocks by detecting the singular vectors "buried" in the word-document matrix. We will partially address this issue here. It is clear that much more work remains to be done.

We can write the document-correlation matrix in the following form:

$$A^T A = M = D + \varepsilon B \tag{7}$$

where $D$ is a decomposable matrix, i.e. a matrix having pure blocks, and $B$ represents spurious correlations between documents. It is important to note that both $D$ and $B$ are symmetric matrices, and this implies that they both have a complete set of orthonormal eigenvectors (and the same happens to $M$).

In general, after performing the SVD of $A$ we get a number of singular values and vectors and only the first one is strictly positive (i.e. all of its entries are >0). It is expected that if there are two blocks and a weak link between them, the associated Perron pair has a set of high positive entries and a complementary set of low values. We also expect that there is another vector where the opposite is true, that is, the coordinates with the lowest values in the first vector now have higher values than the other entries (but in this case there should be negative entries because of the restrictions of theorem 2c, and so we expect the pseudo-zeros to map to one sign and the non-zero entries to map to the other sign). Yet, as we show below, there are more complicated cases.

It is instructive to consider what happens in a miniature eigenvalue problem under perturbation. Consider the following example,

$$D=\begin{pmatrix} a & 0 \\ 0 & b \end{pmatrix}, \quad B=\begin{pmatrix} d & z \\ z & f \end{pmatrix}.$$

The eigenvalues and eigenvectors of D+εB (equation 7) can be calculated as

$$\lambda_1 = a + \varepsilon d + \frac{\varepsilon^2 z^2}{G},  \quad 8\text{i}$$

$$\lambda_2 = b + \varepsilon f - \frac{\varepsilon^2 z^2}{G},  \quad 8\text{ii}$$

where,

$$G = \frac{1}{2}\left[(a+\varepsilon d - b - \varepsilon f) + \sqrt{(a+\varepsilon d - b - \varepsilon f)^2 + 4\varepsilon^2 z^2}\right], \quad 8\text{iii}$$

and,

$$\mathbf{v}_1 = \frac{1}{N}\begin{bmatrix} 1 \\ \frac{\varepsilon z}{G} \end{bmatrix}, \quad 9\text{i}$$

$$\mathbf{v}_2 = \frac{1}{N}\begin{bmatrix} -\frac{\varepsilon z}{G} \\ 1 \end{bmatrix}, \quad 9\text{ii}$$

with N a normalizing factor equal to

$$N = \sqrt{1 + \left(\frac{\varepsilon z}{G}\right)^2}$$

The are three possible cases:

i) If there is no perturbation, ε=0 and the original eigenvalues and eigenvectors of D are recovered. Note that if a=b, then any vector is an eigenvector of eigenvalue *a,* and the basis of orthonormal eigenvectors can be arbitrarily selected.

ii) If the perturbation factor ε≠0 and if εz<< [a-b +ε(d-f)], then the two eigenvectors show a pseudo-zero in one position and a pseudo-one in the other position (since N approaches 1 in this case).

iii) If εz >> [a-b +ε(d-f)] then G≈εz and the absolute value of each component will be equal to 1/sqrt(2).

Notice that case iii) can be reached in a variety of ways. If in fact the connections between blocks are stronger than the blocks themselves case iii) is expected. Yet, it might be the case that εz is small but higher than the relevant difference (a-b +ε(d-f)). An extreme case is given when a=b and d=f (degenerate case), where no matter how small the perturbations are we always expect case iii). This shows that if the first eigenvector has entries of comparative sizes we cannot conclude that the underlying structure is composed by only one block.

In this simple case it can be seen that if there are two "levels" in the magnitudes of entries in the first eigenvector and an opposite tendency in another eigenvector, one can assume that the underlying structure has two (or more) blocks. Nevertheless, there can be cases hard to distinguish from a single irreducible block when the original blocks are similar in size (i.e the Perron-pair for the separate blocks have similar eigenvalues) or are made similar by the perturbation.

To analyze the situation in higher dimensions, we write the eigenvalues and eigenvectors of matrix *M* as perturbation series in ε:

$$\lambda_i^M = \lambda_i + \varepsilon \lambda_i^{(1)} + \varepsilon^2 \lambda_i^{(2)} + \ldots \qquad \text{10i}$$

and,

$$\mathbf{v}_i^M = \mathbf{v}_i + \varepsilon \mathbf{v}_i^{(1)} + \varepsilon^2 \mathbf{v}_i^{(2)} + \ldots , \qquad \text{10ii}$$

where $\lambda_i$ and $\mathbf{v}_i$ are the *i*-th eigenvalue and eigenvector respectively of matrix $\mathbf{D}$ and $\lambda_i^M$ and $\mathbf{v}_i^M$ are the *i*-th eigenvalue and eigenvector of matrix $\mathbf{M}$; the numbers and vectors multiplying the powers of $\varepsilon$ are correcting terms.

Perturbation theory is often used in eigenvalue problems in quantum physics ([Merzbacher 1970]) and has a long tradition in matrix analysis (see [Bellman 1960] and [Stewart 1993]). Strictly speaking, the latter perturbation series (10) are only valid when the eigenvalues are simple but the correction to degenerate cases is easy to obtain.

For non-degenerate cases, the following expression corrects the eigenvalues up to second order in $\varepsilon$ :

$$\lambda_i^M = \lambda_i + \varepsilon \, \mathbf{v}_i^T \mathbf{B} \mathbf{v}_i + \varepsilon^2 \sum_{j \neq i} \frac{|\mathbf{v}_j^T \mathbf{B} \mathbf{v}_i|^2}{(\lambda_i - \lambda_j)} + O(\varepsilon^3) \quad , \qquad 11$$

and to correct the eigenvectors up to first order we write,

$$\mathbf{v}_i^M = \mathbf{v}_i + \varepsilon \sum_{j \neq i} \frac{\mathbf{v}_j^T \mathbf{B} \mathbf{v}_i}{(\lambda_i - \lambda_j)} \mathbf{v}_j + O(\varepsilon^2) \quad . \qquad 12$$

If the characteristic polynomial of $\mathbf{D}$ has a root $\lambda_i$ with multiplicity $g_i > 1$, then the latter equations have to be adapted changing the basis so as to avoid the singularity when $\sigma_i = \sigma_j$.

An illustration of the consequences of equations 11 and 12 can be obtained if the word-document matrix $\mathbf{A}_p$ (where the subscript stands for perturbed $\mathbf{A}$) consist of two blocks of unconnected documents perturbed by the addition of a word used in the two blocks. This word can have a frequent usage within one block and only a sporadic appearance in the other. Alternatively, it might be a generic word used extensively in both blocks (for instance the word "model" in scientific texts). It can also be a polysemous word.

The situation is represented as

$$A_p = \begin{bmatrix} A_1 & 0 \\ \mathbf{w}_1 & \mathbf{w}_2 \\ 0 & A_2 \end{bmatrix} = \begin{bmatrix} A_1 & 0 \\ 0 & 0 \\ 0 & A_2 \end{bmatrix} + \begin{bmatrix} 0 & 0 \\ \mathbf{w}_1 & \mathbf{w}_2 \\ 0 & 0 \end{bmatrix},$$

where $A_1^T A_1$ and $A_2^T A_2$ are symmetric irreducible matrices, representing the correlation of documents within each of the blocks, $\mathbf{w}_1$ and $\mathbf{w}_2$ are file vectors corresponding to the presence of the perturbing word in block 1 ($\mathbf{w}_1$) and its presence in block 2 ($\mathbf{w}_2$). Each of the blocks $A_1$ and $A_2$ have a set of singular values and (right) singular vectors $\{(\sigma_{11}, \mathbf{v}_{11}), (\sigma_{12}, \mathbf{v}_{12}), ...,(\sigma_{1y}, \mathbf{v}_{1y})\}$ $\{(\sigma_{21}, \mathbf{v}_{21}), (\sigma_{22}, \mathbf{v}_{22}), ...,(\sigma_{2\xi}, \mathbf{v}_{2x})\}$.

In terms of document correlations the situation can be described as,

$$M = A_p^T A_p = D + \begin{bmatrix} \mathbf{w}_1^T \mathbf{w}_1 & \mathbf{w}_1^T \mathbf{w}_2 \\ \mathbf{w}_2^T \mathbf{w}_1 & \mathbf{w}_2^T \mathbf{w}_2 \end{bmatrix}$$

where $D = A^T A$ is the matrix of document-document correlations in the unperturbed case and we have set $\varepsilon = 1$. We show below the equations for the perturbed versions of vectors and values corresponding to block 1 ($\mathbf{v}_{1i}^{(M)}, \sigma_{1i}^{(M)}$); the equations for block 2 can be analogously derived. The eigenvalues of $D$ are the the squared singular values of $A_1$ and $A_2$ and its eigenvectors are

$$\mathbf{v}_{1j}^{(D)} = \begin{bmatrix} \mathbf{v}_{1j} \\ \mathbf{0}_r \end{bmatrix}, \text{ and,}$$

$$\mathbf{v}_{2k}^{(D)} = \begin{bmatrix} \mathbf{0}_s \\ \mathbf{v}_{2k} \end{bmatrix}.$$

If we now apply equation 12 we obtain for the perturbed eigenvectors,

$$\mathbf{v}_{1i}^M \approx \mathbf{v}_{1i}^{(D)} + \sum_{j \neq i} \frac{\mathbf{v}_{1j}^{(D)T} B \mathbf{v}_{1i}^{(D)}}{(\lambda_{1i} - \lambda_{1j})} \mathbf{v}_{1j}^{(D)} + \sum_k \frac{\mathbf{v}_{2k}^{(D)T} B \mathbf{v}_{1i}^{(D)}}{(\lambda_{1i} - \lambda_{2k})} \mathbf{v}_{2k}^{(D)}$$

or using the definition of $B$ (and of the singular values),

$$\mathbf{v}_{1i}^{M} \approx \mathbf{v}_{1i}^{(D)} + \sum_{j \neq i} \frac{\langle \mathbf{v}_{1j}, \mathbf{w}_1 \rangle \langle \mathbf{w}_1, \mathbf{v}_{1i} \rangle}{(\sigma^2_{1i} - \sigma^2_{1j})} \mathbf{v}_{1j}^{(D)} + \sum_{k} \frac{\langle \mathbf{v}_{2k}, \mathbf{w}_2 \rangle \langle \mathbf{v}_{1i}, \mathbf{w}_1 \rangle}{(\sigma^2_{1i} - \sigma^2_{2k})} \mathbf{v}_{2k}^{(D)} \qquad (13\ \text{i})$$

and for the singular values,

$$\sigma^2_{i\,(M)} \approx \sigma^2_{i\,(D)} + \langle \mathbf{v}_{1i}, \mathbf{w}_1 \rangle^2 \qquad (13\ \text{ii})$$

This shows that the correction to the eigenvectors include the correlation between the perturbing word and the singular vectors of each block and the difference in the square of the singular values. Although many situations are possible, we sketch some limiting behaviors and exemplify the meaning of equation 13 with our toy example introduced in table 1.

Consider, for instance, what happens to the principal vector if the principal singular values of each block are considerably higher than the other values. In that case only the term corresponding to $k=1$ is included (that is $i=1$, $k=1$ and the other terms can be neglected). The first perturbed vector will be,

$$\mathbf{v}_{11}^{(M)} \approx \frac{1}{N} \begin{bmatrix} \mathbf{v}_{11} \\ \delta \mathbf{v}_{21} \end{bmatrix},$$

where,

$$\delta = \frac{\langle \mathbf{w}_1, \mathbf{v}_{11} \rangle \langle \mathbf{w}_2, \mathbf{v}_{21} \rangle}{\sigma^2_{11} - \sigma^2_{21}}, \quad N = \sqrt{1 + \delta^2}.$$

Notice that when $\sigma^2_{11} \gg \sigma^2_{21}$ the perturbation of the first vector will be seen as positive pseudo zeros, and the second vector (i.e. the one corresponding to the second block) will have negative pseudo-zeros. Both the pseudo-zeros in the first and the second block are of the same magnitude.

In the case of two blocks having similar singular values, perturbations can have a large effect. Provided that the inner products are not too small, if the second singular value approaches the first, then the perturbing block can change dramatically the first singular vector. In order to describe this effect, equations 12 and 13 have to be modified to account for degeneracy, but in general this kind of perturbation leads to a non-negative vector with entries of the same order of magnitude and a complementary vector with mixed signs. This is similar to case iii) discussed after equation 9 and it is related to stability problems in eigenvectors ([Davis and Kahn 1970] see also[Ng et al, 2001a] for a treatment of stability in link analysis). Given that the principal vector of block 2 is symmetrically perturbed, in order to obtain a good representation of the two blocks, both singular vectors should be included, even though the first has no pseudo-zeros and the second has high entries of different signs.

In the case of blocks of dissimilar sizes, many singular values belonging to the first block might be higher that the leading singular value of the second block. The first block will be perturbed by the same block (the first summation in the right hand side of equation 13 i), it will retain non negativity and can have pseudo zeros due to the second block. The situation in $\mathbf{v}_{21}^{(M)}$ can be more complicated, for in general the singular values of each block will have a "heavy tail" of very similar singular values (for instance, this is the case if the first block has the "low-rank-plus-shift" structure studied by Zha and Zhang, see below and [Zha and Zhang 1999]). This necessarily implies that the denominator of perturbing terms in block 2 will be small and the perturbing entries will have different signs. There is no simple way of telling original block entries from perturbations which complicates the construction of a general algorithm.

Besides the differences in singular values, the perturbation effect depends on the scalar products between the perturbing word and the singular vectors of each block. There can be cases where the word is highly frequent in both blocks. If those are the only blocks of documents then this word can be either an unspecific word or a polysemous word. When global weighting schemes are applied, they will receive a very small weight. In these cases (and provided that singular values are well separated) the perturbation will also be seen as pseudo zeros, because the scalar products involved will be very small. But if there are more blocks we have to distinguish a polysemous and a generic word. In those cases where the word is polysemous it has high frequency in both topics, and this might not be affected by global weighting. The detection of the original vectors in this case is only possible when the singular values are separated enough to overcome the scalar product.

In the remaining of this section, we apply these ideas to a perturbed version of Pruned Example 1 (PE1). In section 5 we showed that a minimal truncated version of the matrix *A* of Pruned Example 1 can be successfully obtained keeping only vectors 1, 2 and 5, for these are the only nonnegative vectors. We will illustrate the effect of adding words which connect the perfect blocks, and analyze them in terms of perturbation theory. We reinsert the word *kinetic* to the artificial example of table 1 (PE1), connecting the blocks corresponding to allosteric proteins and Brownian motion, because this word is present in documents 1, 8 and 10. Notice that the meaning of '*kinetic*' in the two contexts is rather different; in the Brownian motion article it refers to the kinetic theory of matter; in the allosteric context it refers to chemical kinetics, as used in mass action law. Thus, this perturbation is an example of polysemy. Due to sparseness we can use as a perturbation of the document-document correlation matrix the 25x25 matrix having zeros except for positions (1,1), (8,8), (10,10), (1,8), (8,1), (1,10), (10,1), (8,10) and (10,8), where it has ones.

In table 3 we show the first thirteen entries of the original singular vectors (corresponding to the first topic -brownian motion- and the second topic -allosteric proteins-) and the perturbing vectors as obtained using equations 13. We only include two vectors from the most important block (documents 6 to 13) and one vector from the smallest block of documents (documents 1 to 5).

|      | V1     | SBP     | DBP    | PV1    |      | V4      | SBP     | DBP    | PV4     |
|------|--------|---------|--------|--------|------|---------|---------|--------|---------|
| d1   | 0      | 0       | 0.0716 | 0.0716 | d1   | 0       | 0       | 0.2351 | 0.2351  |
| d2   | 0      | 0       | 0.0070 | 0.0070 | d2   | 0       | 0       | 0.2090 | 0.209   |
| d3   | 0      | 0       | 0.0062 | 0.0062 | d3   | 0       | 0       | 0.0582 | 0.0582  |
| d4   | 0      | 0       | 0.0009 | 0.0009 | d4   | 0       | 0       | 0.2005 | 0.2005  |
| d5   | 0      | 0       | 0.0009 | 0.0009 | d5   | 0       | 0       | 0.2005 | 0.2005  |
| d6   | 0.3751 | -0.0198 | 0      | 0.3553 | d6   | -0.3012 | -0.0030 | 0      | -0.3042 |
| d7   | 0.1378 | -0.0154 | 0      | 0.1224 | d7   | -0.2071 | -0.0730 | 0      | -0.2747 |
| d8   | 0.3846 | 0.0535  | 0      | 0.4381 | d8   | 0.6176  | 0.0206  | 0      | 0.6382  |
| d9   | 0.2531 | -0.0062 | 0      | 0.2469 | d9   | 0.0248  | 0.0191  | 0      | 0.0439  |
| d10  | 0.3524 | 0.0480  | 0      | 0.4004 | d10  | -0.1693 | 0.1248  | 0      | -0.0445 |
| d11  | 0.2559 | -0.0059 | 0      | 0.2500 | d11  | 0.2194  | -0.0163 | 0      | 0.2031  |
| d12  | 0.5129 | -0.0354 | 0      | 0.4775 | d12  | -0.5082 | -0.0760 | 0      | -0.5482 |
| d13  | 0.4190 | -0.0160 | 0      | 0.4030 | d13  | 0.3864  | 0.1116  | 0      | 0.2748  |

|      | V5     | SBP    | DBP     | PV5     |
|------|--------|--------|---------|---------|
| d1   | 0.1924 | 0.0755 | 0       | 0.2679  |
| d2   | 0.4927 | 0.0086 | 0       | 0.5013  |
| d3   | 0.0503 | 0.0193 | 0       | 0.0696  |
| d4   | 0.5991 | -0.0165| 0       | 0.5826  |
| d5   | 0.5991 | -0.0165| 0       | 0.5826  |
| d6   | 0      | 0      | 0.1200  | 0.1200  |
| d7   | 0      | 0      | 0.0416  | 0.0416  |
| d8   | 0      | 0      | -0.2322 | -0.2322 |
| d9   | 0      | 0      | 0.0040  | 0.0040  |
| d10  | 0      | 0      | 0.1291  | 0.1291  |
| d11  | 0      | 0      | -0.0905 | -0.0905 |
| d12  | 0      | 0      | 0.1616  | 0.1616  |
| d13  | 0      | 0      | -0.2057 | -0.2057 |

SBP= Same-Block Perturbation
DBP= Different-Block Perturbation
V1= First singular vector of PE1.
V4= Fourth singular vector of PE1
V5= Fifth singular vector of PE1

**Table 3**: Singular vectors associated with topic 1 and topic 2, in the original example PE1 (V1, V4 and V5) and after the reintroduction of the word *'kinetic'* (PV1, PV2 and PV5). The original vectors correspond to entries 1 to 13 of the vectors 1, 4 and 5 of table 2. The perturbed versions were calculated using equation 13 (the application of SVD to the full matrix gives slightly different vectors). The same-block perturbation terms are obtained as the summation of all the terms in the expansion that interfere with the same block. Likewise, the different-block perturbation terms are computed adding all terms from the expansion coming from the other blocks. The third topic is not included because it is not changed by the perturbations.

As expected, the first singular vector now has positive coordinates for the two blocks of documents. The contribution to correlation between the two blocks that this first vector makes is small showing the expected pseudo-zeros. Within the first block (documents 1 to 5), the correlation contributed by the first vector is also small, and we need to incorporate the fourth or the fifth vector (or both) to have a significant linkage within the first block.

Whether we have to include only vector 4 or 5 is ambiguous, but notice that both vectors make positive contributions within the first block and mixed contributions to correlation between blocks and within the second block. The fourth vector deteriorates the correlations within the second block more than the fifth. Both vectors make similar contributions to correlation within the first block, suggesting that vector 5, rather than vector 4, should be included. This suggestion is confirmed in figure 6, where we show the cosine matrix using vectors 1, 2 and 5 (fig 6a) and using vectors 1,2, 4 and 5, (fig 6b).

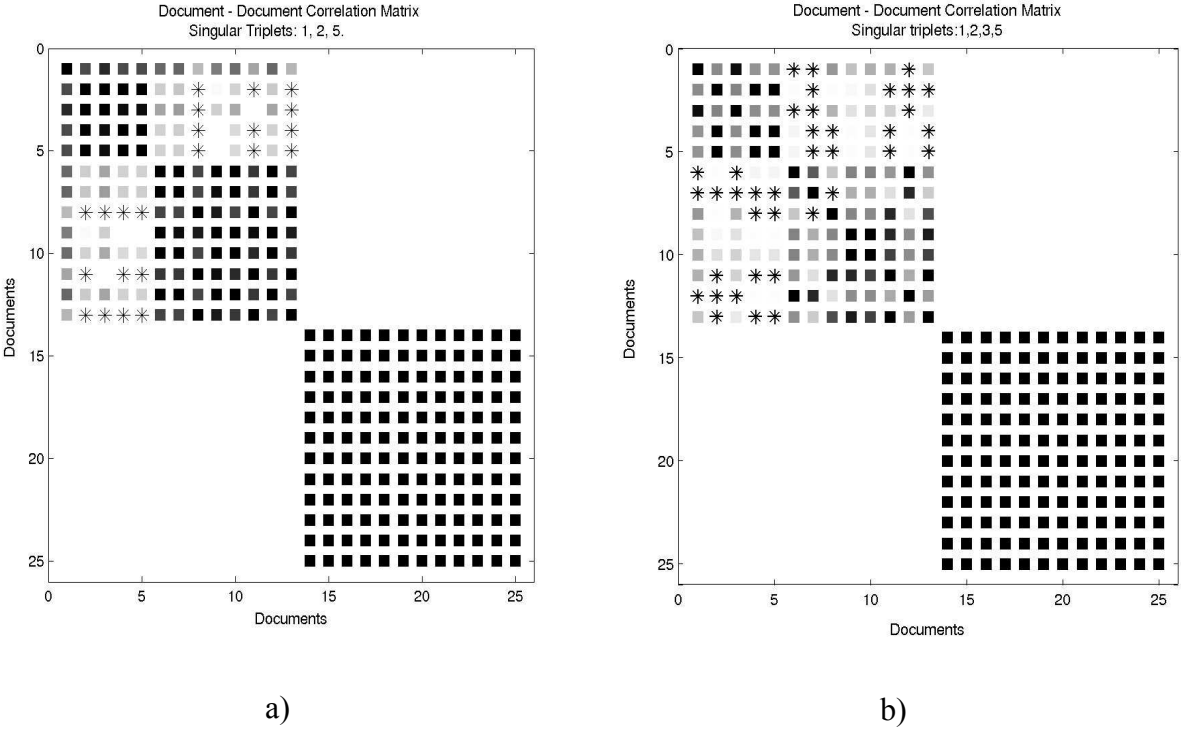

a)                                                          b)

**Figure 6:** Correlation matrices in the perturbed PE1example (compare to table 3) when using the following singular vectors: a) Vectors 1,2 and 5; b) vectors 1,2,4,5. Grayscale= positive cosines greater; ✱ = negative cosines. Notice that in case b) there are negative correlations within the second block and certain correlations are weakened.

# 7. Applications to text collections

The examples we previously used to illustrate the results of our paper are artificial and small, both in number of documents and in number of words. To analyze the applicability of our results to realistic cases we use a subset of the OHSU-MED collection [Hersh et al 1994] as it that we obtained from NIST, (http://trec.nist.gov/data/t9_filtering/filtering.tar.gz ). We randomly selected 10 topics from the subset of queries used by [Hersh & Hickam 1994] which includes 63 topic descriptions and a set of evaluated documents. We included all the evaluated documents that were relevant to these topics, creating a database of 110 documents. We eliminated stop-words and retained 881 words for the analysis. We sorted the documents according to the topic for which they are relevant in order to facilitate comparison. table 4 shows the distribution of documents among topics.

| Topic | Description | # of relevant documents | Documents in matrix |
|---|---|---|---|
| 1 | OHSU13 | 21 | 1-21 |
| 2 | OHSU16 | 13 | 22-34 |
| 3 | OHSU26 | 6 | 35-40 |
| 4 | OHSU29 | 21 | 41-61 |
| 5 | OHSU34 | 5 | 62-66 |
| 6 | OHSU38 | 16 | 67-82 |
| 7 | OHSU57 | 4 | 83-86 |
| 8 | OHSU59 | 10 | 87-96 |
| 9 | OHSU6 | 6 | 97-102 |
| 10 | OHSU8 | 8 | 103-110 |

**Table 4**: Description of the dataset used to test the main results of this paper. Ten topics were selected randomly and all the documents relevant for each topic were retained. In this case all documents are only relevant for one topic (single-topic documents). The data were retrieved from NIST.

---

We performed the singular value decomposition of the full (881 x 110) matrix with and without entropy weighting.

Knowing beforehand which documents are relevant to each topic we detected the leading vectors of each block by two complementary criteria. In the first place, as a rough description of how well each topic is represented by each singular vector, we searched for the vectors that explained most of the square of the Frobenius norm of each topic. We defined the Frobenius norm of each topic as the norm of the submatrix of all documents relevant to the topic, that is,

$$\|T\|_F^2 = \sum_{d_i \in T} \|d_i\|^2 = \sum_{j=1}^{r} \sum_{\substack{i \\ d_i \in T}} \sigma_j^2 v_j^2(i) \quad , \tag{14}$$

where $d_i$ are the documents relevant to topic T and r is the rank of the word-document matrix. The last equality follows directly from equation (6). The contribution to the square of the norm that each singular vector makes is given by the summation over *i* in equation (14). table 5 shows the most representative singular vectors for each topic according to the Frobenius norm.

| TOPIC | 1st vector | 2nd vector | 3rd vector | 4th vector | Fraction of $\|T\|_F^2$ |
|---|---|---|---|---|---|
| 1 | 3 | 1 | 22 | 7 | 0.3331 |
| 2 | 1 | 2 | 7 | 10 | 0.5642 |
| 3 | 5 | 4 | 6 | 1 | 0.4898 |
| 4 | 1 | 2 | 4 | 5 | 0.4693 |
| 5 | 24 | 26 | 1 | 58 | 0.1961 |
| 6 | 1 | 9 | 15 | 12 | 0.2897 |
| 7 | 45 | 17 | 25 | 26 | 0.3486 |
| 8 | 8 | 16 | 1 | 6 | 0.4236 |
| 9 | 1 | 11 | 3 | 25 | 0.4270 |
| 10 | 12 | 1 | 28 | 9 | 0.3628 |

**Table 5**: The most important singular vectors for each topic in terms of the square of Frobenius norm. For each topic, we calculated the percentage of the total norm that is accounted for by each vector and then sorted them in descending order (from left to right) of percentage. We show the first four vectors for each topic. In the fifth column we show the percentage of the square of the topic norm that is captured by these four vectors (and values) Notice that some topics' norms are highly retained by just 3 or 4 vectors (for instance topics 2 and 3) while others are distributed over many vectors (for instance, see topics 1, 5 and 6).

---

The other criterion was based on performing the dot product between a vector having ones only in positions corresponding to the relevant documents (for one topic) and the columns of V. We retained those columns that yielded the three highest absolute dot products (for each topic) and then inspected each column to select those that more closely represent one topic. Using these procedures we selected the following singular vectors approximately representing each topic (table 6).

| Right Singular vector | Topic |
|:---:|:---:|
| 1 | 2 |
| 2 | 4 |
| 3 | 1 |
| 5 | 3 |
| 8 | 8 |
| 9 | 6 |
| 11 | 9 |
| 12 | 10 |
| 17 | 7 |
| 26 | 5 |

**Table 6**: Leading vectors of each block selected by inspection of the vectors obtained by applying the dot-product criterion (see text). Topics 2 and 4 are both highly represented in the first and second singular vectors (see table 5). Notice that the vectors that we selected are among the two more representative of the norm of each topic. In most cases the the dot-product criterion leads correctly to the leading vectors, except for topic 7, where the dot product is higher with singular vector 25, but due to the fact that some documents relevant to topic 7 are overrepresented and others underrepresented in vector 25, we selected vector 17.

---

With these vectors we constructed an approximation to the word-document matrix. To test whether this matrix retains the topic structure we compared the precision-recall curves in the following cases:

1. The matrix of leading vectors (rank =10) obtained by the procedure outlined.

2. A matrix obtained by retaining the first 18 singular values (which is the best LSA approximation).

3. A matrix obtained by performing SVD after entropy weighting, and retaining the first 18 singular values.

4. A matrix obtained by retaining the first ten singular values.

We used the relevance judgements and queries that are distributed with the OHSU-MED database. In figure 7 we compare precision-recall curves in the four situations.

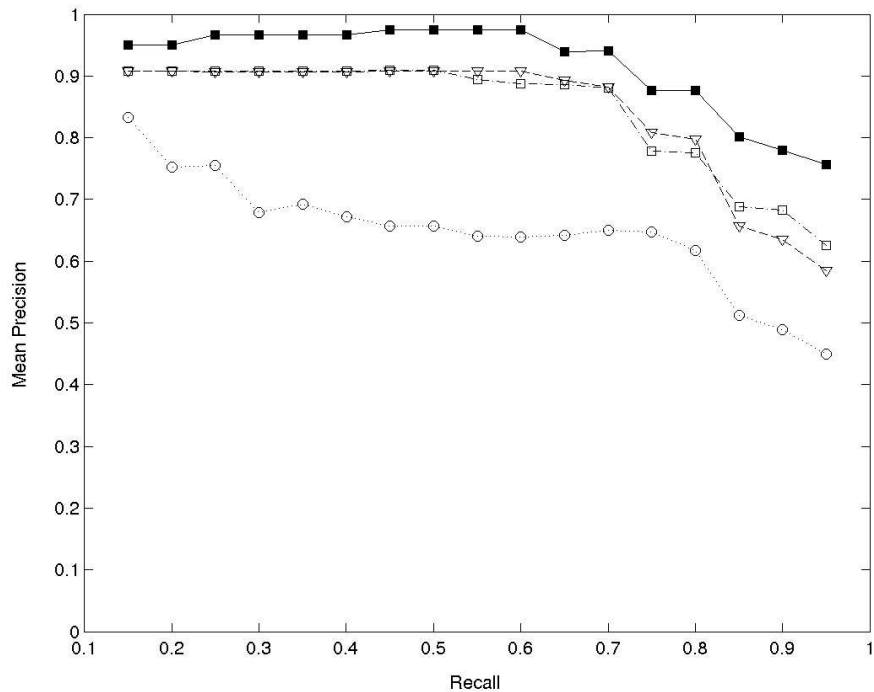

**Figure 7**: Mean precision vs recall of the different cases described in text: (–■–) case 1, the approximation was composed by the 10 vectors (and values) selected in table 6;(- ▽ -) case 2, the svd best rank -18 approximation; (· ·□–) case 3, the best rank-18 approximation after entropy weighting; (· ○ ·) case 4, the best rank-10 approximation. The mean precision for each level of recall was calculated as the mean over all the relevance judgements. Although the relevance judgements are expressed in a three-valued scale (0 irrelevant, 1 probably relevant, 2 definitely relevant) we considered relevant for one topic all documents that are at least probably relevant for this topic.

The mean precision over the different levels of recall is given in table 7.

|  | case 1 | case 2 | case 3 | case 4 |
|---|---|---|---|---|
| mean precision: | 0.9200 | 0.8432 | 0.8448 | 0.7405 |
| % variation in relevance (100% =case 2): | 9.11 | 0 | 0.19 | -12.18 |

**Table 7**: Mean precision for the four cases considered. The mean precision was calculated in the four cases averaging over all recall levels (17 levels from 0.15 to 0.95).

Although compared to case 2, the improvement in case 1 is of only 9 %, it is remarkable that this performance can be obtained retaining only ten singular vectors. It has to be noted that case 2 shows a very high mean precision, leaving little room for improvements. Indeed, the full word-document matrix performs almost as well as case 2 (mean precision 0.8346). Nevertheless the precision was higher for all levels of recall in case 1 compared to cases 2 and 3. Moreover, compared to case 4, which is the best rank-10 approximation our rank-10 matrix performs far better. Although the matrix used in case 4 has as many factors as there are topics, it performs relatively poorly in this task. Thus, as we showed for example PE1, retaining the first k vectors from the SVD is not enough to capture the k-topic structure. The conclusion from this experiment confirms one of the main results of this paper, namely, that a better description of a text-database can be obtained by omitting some singular values and vectors, and retaining the leading vectors for each topic.

The other main result, which shows that the leading vectors have a perturbed Perron-like structure is much more difficult to apply, although it should help in selecting the relevant vectors when no information about relevance is known. According to the previous sections, in an ideal case, each topic has an associated set of singular vectors, and only the first of each topic (the leading vector) is non-negative. Thus, one can find the set of vectors corresponding to each topic by searching non-negative vectors and those that share with them the non-zero entries. In a perturbed case, in general the only nonnegative vector will be the first and the zeros are replaced by perturbing terms. If the perturbing terms are small the other leading vectors should have large entries of one sign and small entries of mixed signs. In fact, the idea to use a dot product in order to detect the leading vectors that we used to construct case 1 was inspired by the notion that leading vectors should have high entries corresponding to one topic and small numbers in the other positions. Vectors having ones only in positions corresponding to relevant documents for one topic yield higher dot products with leading vectors than with non-leading vectors, at least as far as perturbations are small.

Perturbations can have an important effect though, and in many cases equations 12 and 13 do not apply, because the numerators in the perturbing terms can be higher than the difference in the square of the singular values. For instance, topics 2 and 4 in the OHSU-MED example, although different from each other, coexist in the first two singular vectors, mainly due to the presence of generic words that appear repeatedly in both topics, like "PATIENS" which is the most frequent (and most entropic) word in the database. Notice that according to table 5, topic 2 is preferentially represented by the first and second vector, whereas topic 4 is preferentially represented by the second and first vector. Eliminating the word "PATIENTS" partially separates the two topics. If this is accomplished, the first vector corresponds mainly to topic 2 and the second vector contains the fourth topic. Moreover these two vectors display the pattern of pseudo-zeros anticipated for perturbed Perron vectors (data not shown). Nevertheless, the elimination of the most entropic words distorts the representation of other topics and seems to be a bad strategy for discovering the leading vectors if it is applied blindly.

In spite of these difficulties we think that a procedure to discover the leading vectors can be devised, if particular care is taken to treat large perturbations and small perturbations separately. This is a possibility that has to be explored further.

## 8. Related work

In the last few years several contributions to the understanding of LSA have been published. [Papadimitriou et al 2000; Ando and Lee 2001; Azar et al, 2001]. These authors use invariant subspace perturbation theory to derive conditions under which LSA works properly.

In particular, for a particular set of hypothesis, Papadimitriou and coworkers have shown that provided that the number of factors included in the truncated SVD matches the number of underlying topics and the documents are large enough, LSA will discover these topics. Their results are applicable in those special cases in which the term distribution is not altered by style, each document is on a single topic, and the terms are partitioned among the topics so that each topic distribution has high probability on its own terms and low probability on all others.

These conditions are similar to our perturbed topics, yet there is an important difference. In the work of Papadimitriou and co-workers, the first k singular vectors capture the k topics, whereas in our formulation there is no guarantee that the first k values correspond to k topics.

Ando and Lee [2001] have shown that the working of LSA depends on the uniformity of the underlying topic structure. They argue that the topic model of Papadimitriou and coworkers implies that the topic distribution is close to uniform. In particular, this is required for the second singular value of each of the blocks to be smaller than any of the principal singular values of the other blocks. The problems of LSA with non-uniform topic distributions are related to one of the main points of this paper, namely, that in order to represent k topics whenever they are of different importance, more than k successive singular values and vectors have to be included, or alternatively some singular values should be omitted.

Ando and Lee's analysis of LSA does not require single-topic documents, nor particular word distributions. They have also proposed an algorithm, Iterative Residual Rescaling (IRR), to deal with the non-uniform case [Ando 2000; Ando and Lee 2001]. To understand IRR, consider that in SVD the left singular vector $\mathbf{u}_j$ points in a direction which is a compromise between the direction of the majority of residual vectors and the longer residual vectors. These residuals are obtained after the projection of each document onto the span of $\mathbf{u}_1, ..., \mathbf{u}_{j-1}$. By weighting residuals as a power of their norms, IRR enhances the longer residuals corresponding to under-represented topics over the possibly high number of short residuals coming from leading topics.

It is interesting then, to compare the output of IRR for Pruned Example 1 and the singular vectors. In order to achieve this goal we need to estimate a scaling factor. Given that we are dealing in PE1 with single topic documents we can directly calculate the measure of non-uniformness suggested by Ando & Lee. Yet the best results are given for a different scaling factor (see table 8 for details).

| a) | | | b) | | |
|---|---|---|---|---|---|
| $B_1$ | $B_2$ | $B_3$ | $u_1$ | $u_2$ | $u_5$ |
| 0.7002 | 0 | 0 | 0.7226 | 0 | 0 |
| 0.3295 | 0 | 0 | 0.3028 | 0 | 0 |
| 0 | 0 | 0.7713 | 0 | 0 | 0.7699 |
| 0.2376 | 0 | 0 | 0.2275 | 0 | 0 |
| 0.2376 | 0 | 0 | 0.2275 | 0 | 0 |
| 0 | 0.3454 | 0 | 0 | 0.3423 | 0 |
| 0 | 0.1456 | 0 | 0 | 0.1613 | 0 |
| 0.1656 | 0 | 0 | 0.1813 | 0 | 0 |
| 0 | 0.3233 | 0 | 0 | 0.3046 | 0 |
| 0.2039 | 0 | 0 | 0.1842 | 0 | 0 |
| 0 | 0.4326 | 0 | 0 | 0.4566 | 0 |
| 0 | 0 | 0.1005 | 0 | 0 | 0.1105 |
| 0 | 0 | 0.5471 | 0 | 0 | 0.5456 |
| 0 | 0 | 0.3092 | 0 | 0 | 0.3120 |
| 0.2686 | 0 | 0 | 0.2513 | 0 | 0 |
| 0 | 0.2924 | 0 | 0 | 0.2862 | 0 |
| 0 | 0.6744 | 0 | 0 | 0.6678 | 0 |
| 0 | 0.1442 | 0 | 0 | 0.1367 | 0 |
| 0.1404 | 0 | 0 | 0.1714 | 0 | 0 |
| 0 | 0.0823 | 0 | 0 | 0.0947 | 0 |
| 0.3570 | 0 | 0 | 0.3511 | 0 | 0 |

**Table 8.** The output of IRR and SVD for PE1. $BB^T$ is analogous to $U_k U_k^T$ in SVD a) Matrix $B$, obtained by IRR restricted to a 3 dimensional space and scaling factor equal to 0.4305. Using lower scaling factors eliminates the Brownian motion block. Higher factors can eliminate the third block. b) The first, second and fifth left singular vector of PE1. The differences in each entry are small. The shaded entries correspond to the most important relative differences. In this case they are not higher than a 3.5 %.

The columns of matrix B are very similar to left singular vectors 1, 2 and 5 (taking the left singular vectors facilitates comparison). So in this simple case, where we know beforehand the number of topics and have a simple way of estimating the scaling factors, IRR discovers topic vectors which are not too far from the leading singular vectors of each block.

There are some differences between the approach behind IRR and searching for the Perron vectors of each block. In particular, we propose that provided that the "buried" leading vectors can be detected, the dimensions selection problem can be solved by selecting just as many as these vectors. Ando and Lee [Ando 2000; Ando and Lee 2001] also discuss other approaches to solving the dimensions selection problem, like a log-likelihood method or a training dataset.

Another difference we did not explore further is that, in order to retrieve minor topic documents, IRR weights selectively the residuals. This can change the basis vectors with respect to SVD vectors, modifying the relations between words. Thus, despite similarities between Ando & Lee's approach and ours, the idea that LSA works well when it is able to discover the Perron vectors of each topic is, from our point of view, an interesting and different way of formulating the problem.

Zha and Zhang [Zha & Zhang, 2000] have discussed the consequences of word-document matrices having an approximate low-rank-plus-shift structure and they have shown that in these cases LSA works well. The existence of this structure is particularly relevant since it allows for the use of simple updating and partial algorithms to compute the SVD. We did not explore the connections between this approach and ours.

In some respect, our approach has ties to Nonnegative Matrix Factorization (NMF) [Lee and Seung, 2000] a technique aimed at representing data as a product of a sparse nonnegative matrix of features and a nonnegative matrix of coefficients. The virtue of this approach is that the elements of the feature base usually have a clear interpretation. For instance, the approximate algorithms of Lee and Seung can learn a parts-based representation of faces.

Moreover it has been shown that when applied to a database of documents, the method groups semantically related words, solving polysemy and synonymy. In the simplified example PE1 (where there are clearly separated blocks of documents) choosing the first, the second and the fifth singular values and their associated vectors is equivalent to performing a nonnegative matrix factorization. Nevertheless, there is no guarantee that this factorization optimizes the usual cost functions in NMF (as the Frobenius norm of the difference between the original matrix and the approximation). In table 9 we compare the output of five-dimensional NMF and a matrix built on the three vectors we used to represent PE1.

|      | $H^T(1)$ | $H^T(2)$ | $H^T(3)$ | $H^T(4)$ | $H^T(5)$ | V1 | V2 | V5 |
|------|----------|----------|----------|----------|----------|--------|--------|--------|
| d1   | 0 | 0 | 0 | 0.1745 | 0 | 0 | 0 | 0.1924 |
| d2   | 0 | 0 | 0 | 0.4910 | 0 | 0 | 0 | 0.4923 |
| d3   | 0 | 0 | 0 | 0 | 0 | 0 | 0 | 0.0503 |
| d4   | 0 | 0 | 0 | 0.6035 | 0 | 0 | 0 | 0.5991 |
| d5   | 0 | 0 | 0 | 0.6035 | 0 | 0 | 0 | 0.5991 |
| d6   | 0 | 0 | 0.5209 | 0 | 0 | 0.3751 | 0 | 0 |
| d7   | 0.0922 | 0 | 0 | 0 | 0 | 0.1379 | 0 | 0 |
| d8   | 0.6861 | 0 | 0 | 0 | 0 | 0.3846 | 0 | 0 |
| d9   | 0.1038 | 0 | 0.2461 | 0 | 0 | 0.2531 | 0 | 0 |
| d10  | 0 | 0 | 0.4692 | 0 | 0 | 0.3524 | 0 | 0 |
| d11  | 0.2926 | 0 | 0.0973 | 0 | 0 | 0.2559 | 0 | 0 |
| d12  | 0 | 0 | 0.6622 | 0 | 0 | 0.5129 | 0 | 0 |
| d13  | 0.6515 | 0 | 0 | 0 | 0 | 0.4190 | 0 | 0 |
| d14  | 0 | 0 | 0 | 0 | 0.2574 | 0 | 0.0805 | 0 |
| d15  | 0 | 0 | 0 | 0 | 0.3426 | 0 | 0.1588 | 0 |
| d16  | 0 | 0.3659 | 0 | 0 | 0.3494 | 0 | 0.4907 | 0 |
| d17  | 0 | 0.0223 | 0 | 0 | 0.5909 | 0 | 0.3123 | 0 |
| d18  | 0 | 0.4657 | 0 | 0 | 0 | 0 | 0.3514 | 0 |
| d19  | 0 | 0.3814 | 0 | 0 | 0 | 0 | 0.2884 | 0 |
| d20  | 0 | 0.3918 | 0 | 0 | 0 | 0 | 0.3319 | 0 |
| d21  | 0 | 0 | 0 | 0 | 0.3426 | 0 | 0.1588 | 0 |
| d22  | 0 | 0.0587 | 0 | 0 | 0.1055 | 0 | 0.1060 | 0 |
| d23  | 0 | 0 | 0 | 0 | 0.4606 | 0 | 0.1918 | 0 |
| d24  | 0 | 0.2613 | 0 | 0 | 0 | 0 | 0.1752 | 0 |
| d25  | 0 | 0.5273 | 0 | 0 | 0.0670 | 0 | 0.4574 | 0 |
|      | a) | | | | | b) | | |

**Table 9**: a) Matrix $H^T$ obtained by the Nonnegative Matrix Factorization (NMF) of the word-document matrix corresponding to PE1, $A=WH$, by restricting the dimensions of matrix $H$ to be 5x25. b) The first, second and fifth right singular vector of matrix $A$. Nonnegative matrix factorization was computed using the package nmfpack of Patrik Hoyer (www.cs.helsinki.fi/patrik.hoyer, see [Hoyer, 2004]).

As can be seen, each of the output vectors derived by NMF and each of the leading vectors calculated using SVD correspond to only one topic. In NMF the relevant documents are distributed among more vectors than in SVD, and document 3 is absent in the output of NMF. Reducing the dimensions of NMF to three leads to the elimination of the "Brownian motion" topic. Thus, in spite of the reliance of both approaches on nonnegative vectors, the outputs show important differences. Moreover, when using NMF in the perturbed version of table 3 intratopic correlations are deteriorated, at least when using five dimensions (data not shown). Of course, in perturbed cases there is in general only one non-negative singular vector and the rest have mixed signs.

## 9. Perspectives and Concluding remarks

Initially used as an information retrieval tool, LSA has been applied to tackle different but related problems like text categorization and question answering [Dumais 2003]. We believe that the strength and wide applicability of the method lies in its ability to discover hidden topics. Under the cluster hypothesis, closely associated documents tend to be relevant to the same queries [van Rijsbergen 1976]. Thus, finding the underlying topics should enhance performance in retrieval. At the same time, the separation of the different contexts (i.e. topics) where the multiple meanings of polysemous words are expressed, allows for the disambiguation of these words. Likewise, synonyms are recognized as such because they tend to occur within the same topics (but compare to [Higgins 2004]). Finally, some aspects of human memory are adequately modeled by LSA precisely by the same reason: by finding topics, LSA groups together similar objects.

LSA is not the only available technique for latent structure detection. Essentially various partitioning and indexing techniques are aimed at the same objective. Then, why do we concentrate on LSA? First of all, it is one of the few techniques that has been compared to humans in its memory performance [Landauer & Dumais 1997]. Second, its mathematical formulation has close ties to linear neural networks (see for instance [Pomi & Mizraji 1999]) rendering it easy to implement in a connectionist model. We hope to use this relation to build a connectionist model of text analysis. Third, since it is based on spectral techniques, LSA is founded on well known theorems of linear algebra as those applied to multivariate data analysis and information retrieval (for a review on the application of linear algebra to information retrieval see [Langville and Meyer 2004]).

In this work we focused upon trying to understand how LSA detects the underlying topic structure in a text database. In spite of the theoretical foundations of the mathematical tools used in the application of Latent Semantic Analysis, the reasons that make the method work are not so clear. We discussed several studies aimed at a theoretical explanation of LSA's merits and weaknesses (section 8). The lack of a general theory is evinced when trying to apply LSA to a particular database. For instance, there is no universal recipe to determine the appropriate number of dimensions, although several criteria have been applied (for instance, see [Ando 2000; Azar et al 2001]). If the multivariate analysis literature is to be followed one should apply Cattel's scree test [Cattel 1966], searching for a noticeable drop off in the profile of singular values. Nevertheless, centering the attention on the structure of singular vectors might be more fruitful than only considering the profile of singular values. The are two reasons for this approach. In the first place, as we show above, the structure of the singular vectors contains the relevant information of linkage between documents in a transparent manner. In the second place, in those cases where each topic is differentially represented in the database, in order to represent all the topics it is necessary either to include vectors that deteriorate within-topic correlations or to exclude some vectors of the more important topics, retaining only a small amount from each block. We have shown that in particular cases the vectors to be retained are marked by a particular structure.

The main point in our work is that we argue that the topic structure can be thought of as a perturbed block structure in the document-document correlation matrix (or likewise in the word-word correlation matrix). Provided the Perron pairs belonging to each block can be detected, the topics can be represented just by retaining as many vectors as Perron vectors there are. Notice that in this case the dimensions selection problem can be automatically solved. The procedure of selecting the first k singular vectors to represent k topics is only appropriate when the underlying blocks are of similar sizes (i.e. the topics are similarly represented in the database by documents of comparable sizes). If this is not the case, more singular vectors must be included in order to increase recall, but at the expense of deteriorating precision. It is common in applications of LSA (see as an example figure 5 in [Landauer et al. 1998] or figure 1 in [Dumais 1991]) to find more than one relative maximum in the performance (as a function of dimensions) of the reconstructed database, a fact that we believe depends on the necessity to include more factors than topics to reach all the topics. Another strategy would be to include only the leading vectors of each block. Although we did not prove that performance is maximized when only the leading vectors are selected, we showed that excluding some singular vectors can improve performance in a subset of OHSU-MED.

These considerations are easy to apply when the word-document matrix is decomposable, a fact that can only happen when there are single topic documents and no polysemous terms. Even generic terms that are not eliminated in preprocessing contribute to the distortion of the decomposable character. We showed some examples of how the leading vectors are distorted by perturbations. It has long been recognized that eigenvectors can be particularly unstable under perturbations, and that invariant subspaces are robust [Davis & Kahan 1970]. Ng and coworkers have shown that small perturbations can have a tremendous effect in link analysis [Ng et al 2001a; 2001b]. We showed that perturbations by a single word can have an important effect in singular vectors when the difference in the square of the singular values of each block is small (equations 13). In those cases, the original leading vectors are significantly affected, and the perturbing terms are not small "pseudo-zeros". Both vectors (or more if there are more blocks with similar singular values) should then be retained, in line with arguments from invariant subspace perturbation theory [Ng et al 2001a]. If we know beforehand what terms are perturbing (as in our simple example) then it is easy to see the resulting problem as a perturbation situation. Yet, when trying to apply these concepts to information retrieval or cognitive function we are confronted with an inverse problem, i.e. deducing from the singular value decomposition of a perturbed matrix an underlying topic structure.

As everything that has been said about the documents apply as well to the words, Perron-Frobenius theory of nonnegative matrices has shown to be valuable to understand one of the strengths of LSA, namely, its ability to deal with synonyms and the documents that share them. If two documents are linked by a chain of words, then the first singular vector will group them together, as Lemma 1 shows, even if they do not share any word. Two words in that context will be identified as synonyms. It is clear that this alone would make the method prone to problems of polysemy. A polysemous word shared by various unrelated documents would cause the documents to be correlated. In our examples though, the dot product between distantly related documents is small when one singular vector is retained for each topic. Moreover, although we did not explore the issue, we believe that some of the weighting schemes [Dumais 1991], notably weighting by 1-entropy, reduce the perturbation introduced by polysemous words, at least of those that are overrepresented.

Perron-Frobenius Theory and Perturbation Theory give us a set of interesting tools. We hope that working through these lines can yield interesting results that help us understand the powers and limitations of LSA.

## 10. Acknowledgments

We would like to thank MSc. Laura Quintana for her comments. We thank the anonymous referees of an earlier version of this manuscript for many helpful suggestions. This work was partially supported by PEDECIBA (Uruguay).

# 12. References

Ando, R.K. Latent semantic space: Iterative Scaling improves inter-document similarity measurement. Proceedings of the 23rd SIGIR, (2000), 216-223.

Ando R. K., and Lee L. Iterative Residual Rescaling: An analysis and generalization of LSI. Proceedings of the 24th SIGIR, (2001), 154-162.

Azar, Y., Fiat, A., Karlin, A., McSherry, F. and Saia, J. Spectral analysis of data. Proceedings of the ACM Symposium on Theory of Computing (STOC), 2001, 619-626.

Bellman, R. Introduction to Matrix Analysis. McGraw-Hill, New York, (1960).

Berry, M., Dumais, S. and O'Brien, G. Using Linear algebra for intelligent information retrieval. SIAM Review, 37, 4, (1995), 573-595.

Cattell, R. B. The scree test for the number of factors. Multivariate Behavioral Research, 1, (1966), 629-637.

Chomsky, N. Knowledge of Language: its Nature, Origins and Use. Praeger Publishers, New York, (1985).

Davis, C. and Kahan, W.M. The rotation of eigenvectors by a perturbation III. SIAM Journal on Numerical Analysis, 7, (1970), 1-46.

Deerwester, S, Dumais, S, Furnas, G, Landauer, T and Harshman, R. Indexing by latent semantic analysis. Journal of the American Society for Information Science. 41,(1990),391-407.

Dumais, S. Improving the retrieval of information from external sources. Behavior Research Methods, Instruments and Computers, 23, 2, (1991), 229-236.

Dumais, S. Data-driven approaches to information access. Cognitive Science, 27, (2003), 491-524.

Hersh, W, Buckley, C, Leone, TJ, Hickam, D. Ohsumed: An interactive retrieval evaluation and new large test collection for research. Proceedings of the 17th annual ACM SIGIR Conference, (1994), 129-2001.

Hersh, W, Hickam, D. Use of a multi-application computer workstation in clinical settings. Bulletin of the Medical Library Association, (1994), 82: 382-389.

Higgins, D Which statistics reflect semantics? Rethinking synonymy and word similarity. In "International Conference on Linguistic Evidence". Tübingen, January 29 - 31, (2004) (available on line at: http://www.sfb441.uni-tuebingen.de/ling.evidence/abstracts/higgins.pdf)

Hoffmann., T. Probabilistic latent semantic indexing. In Proceedings on the 22nd annual international ACM SIGIR conference, (1999), 50-57.

Hoyer, PD (2004) Non-negative matrix factorization with sparseness constraints. Journal of Machine Learning Research, 5, (2004), 1457-1469.

Landauer, T. and Dumais, S. A solution to Plato's problem: The latent semantic analysis theory of acquisition, induction and representation of knowledge. Psychological Review, 104, 2, (1997), 211-240.

Landauer, T., Foltz, P.W. and Laham, D. An introduction to latent semantic analysis. Discourse Processes, 25, 2, (1998), 259-284.

Langville, A.N. and Meyer, C.D. A survey of eigenvector methods for web information retrieval. (to appear in The SIAM Review) available on line at: http://meyer.math.ncsu.edu/Meyer/PS_Files/Survey.pdf (2004).

Lee, D.D and Seung, H.S. Learning the parts of objects by non-negative matrix factorization. Nature, 401, (1999) 788,791.

Lewis, J.D. and Elman, J.L. Learnability and the Statistical Structure of Language: Poverty of Stimulus Arguments Revisited. In Proceedings of the 26th Annual Boston University Conference on Language Development 1, (2002), 359-370.

Meyer C. D. Matrix Analysis and Applied Linear Algebra. SIAM, Philadelphia, (2000).

Merzbacher, E. Quantum Mechanics. Second Edition. Wiley & Sons, New York, (1970).

Montemurro, M.A.and Zanette, D.H. Entropic analysis of the role of words in literary texts. Advances in Complex Systems, 5, 1, (2002), 7-17.

Ng, A.Y., Zheng, A.X., Jordan, M.I. Link analysis, eigenvectors and stability. Proceedings of the 17th International Joint Conference on Artificial Intelligence. (2001a), 903-910.

Ng, A.Y., Zheng, A.X., Jordan, M.I. Stable algorithms for link analysis. Proceedings of the 24th Annual International ACM SIGIR Conference, (2001b), 258-266.

Papadimitriou, C. H., Raghavan, P., Tamaki, H. and Vempala, S. Latent semantic indexing: a probabilistic analysis. Journal of Computer and System Sciences. 61, (2000), 217-235.

Pomi, A. and Mizraji, E. Memories in context. Biosystems. 50, (1999), 173-188.

Stewart, G.W. On the early history of the singular value decomposition. SIAM Review 35,4, (1993), 551-566.

van Rijsbergen, C.J. Information Retrieval. London: Butterworths, (1979).

Zha, H. and Zhang, Z. On matrices with low-rank-plus shift structure: partial svd and latent semantic indexing.SIAM Journal on Matrix Analysis and Applications, 21 (1999), 522-536.